\documentclass[twocolumn,pra,showpacs,superscriptaddress]{revtex4-1}
\usepackage{graphicx}
\usepackage{amssymb,amsmath}
\usepackage{bm}% bold math
\usepackage{dcolumn}
\usepackage{subfigure}
\usepackage{url}

\begin{document}

\title{Tunable spin-orbit coupling synthesized with a modulating gradient magnetic field}

\author{Xinyu Luo}
\affiliation{State Key Laboratory of Low Dimensional Quantum
Physics, Department of Physics, Tsinghua University, Beijing 100084,
China}

\author{Lingna Wu}
\affiliation{State Key Laboratory of Low Dimensional Quantum
Physics, Department of Physics, Tsinghua University, Beijing 100084,
China}

\author{Jiyao Chen}
\affiliation{State Key Laboratory of Low Dimensional Quantum
Physics, Department of Physics, Tsinghua University, Beijing 100084,
China}

\author{Qing Guan}
\affiliation{Beijing National Laboratory for Condensed Matter Physics, Institute of Physics, Chinese Academy of Sciences,
Beijing 100080, China.}

\author{Kuiyi Gao}
\affiliation{Beijing National Laboratory for Condensed Matter Physics, Institute of Physics, Chinese Academy of Sciences,
Beijing 100080, China.}

\author{Zhi-Fang Xu}
\affiliation{Department of Physics and Astronomy, University of Pittsburgh, Pittsburgh, Pennsylvania 15260, USA}

\author{L. You}
\affiliation{State Key Laboratory of Low Dimensional Quantum
Physics, Department of Physics, Tsinghua University, Beijing 100084,
China} \affiliation{Collaborative Innovation Center of Quantum
Matter, Beijing, China}

\author{Ruquan Wang}
\email{ruquanwang@iphy.ac.cn}
\affiliation{Beijing National Laboratory for Condensed Matter Physics, Institute of Physics, Chinese Academy of Sciences, Beijing 100080, China.} \affiliation{Collaborative Innovation Center
of Quantum Matter, Beijing, China}

\date{\today}% It is always \today, today,

\begin{abstract}
We report the observation of tunable spin-orbit coupling (SOC) for ultracold $^{87}$Rb atoms in
hyperfine spin-1 states.
Different from most earlier experiments where atomic SOC of pseudo-spin-1/2
are synthesized
with Raman coupling lasers, the scheme we demonstrate employs a gradient magnetic
field (GMF) with ground state atoms and is immune to atomic spontaneous emission.
The effect of the SOC is confirmed through the studies of:
1) the collective dipole oscillation of an atomic condensate in a harmonic trap
after the synthesized SOC
is abruptly turned on; and 2) the minimum energy state at a finite adiabatically
adjusted momentum when the SOC strength
is slowly ramped up. The coherence properties of the spinor condensates
remain very good after interacting with modulating GMFs,
which prompts the enthusiastic claim that our work provides a new repertoire
for synthesized gauge fields aimed at quantum simulation studies with cold atoms.
\end{abstract}


\maketitle

Spin-orbit coupling (SOC), as is often referred to in condensed matter physics, couples the spin of a particle to its orbital degrees of freedom. It is believed that SOC constitutes an important
enabling element for quantum simulation studies with ultracold atoms~\cite{DalibardRMP2011,SpielmanReview2013}.
Research on SOC is an active area of study due to its ubiquitous
appearance in condensed matter phenomena, such as topological insulator \cite{KaneRMP2010,QiRMP2011}, spin Hall effect \cite{KatoScience2004,KonigScience2007}, and spintronics \cite{ZuticRMP2004}. In contrast to solid-state materials, where SOC originates from the orbital motion of electrons inside a crystal's intrinsic electric field, neutral atoms interact with electromagnetic fields
differently, and thus atomic SOC is often synthesized from spin-dependent gauge fields.
Recent years have witnessed significant successes in this direction \cite{SpielmanNat2009,SpielmanNat2011,StruckPRL2012,JimenezGarcPRL2012, PanPRL2012,ZhangPRL2012,MartinPRL2012,AidelsburgerPRL2013,MiyakePRL2013,JotzuNat2014}.
A popular scheme employs Raman laser fields~\cite{SpielmanNat2011} coupled to
two atomic ground states forming a pseudo-spin-1/2 system to synthesize a SOC with equal Rashba \cite{RashbaJPC1984} and Dresselhaus \cite{DresselhausPR1955} contributions.
This is routinely used nowadays for both bosonic \cite{SpielmanNat2011,PanPRL2012} and fermionic \cite{ZhangPRL2012,MartinPRL2012} alkali atom species. More general forms of synthetic gauge fields are pursued actively in a variety of settings,
which together with the above well understood Raman scheme and the highly controllable ultracold atomic systems significantly expand the scopes and the abilities of quantum simulation studies, fostering exciting opportunities for observing novel quantum phenomena with ultracold atoms \cite{BeelerNat2013,ZhangNat2014,PanNat2014}.

The Raman scheme, pioneered by the Spielman group \cite{SpielmanNat2011}, makes use of coherent atom-light interaction. As pointed out by several authors~\cite{SpielmanReview2013,KennedyPRL2013}, spontaneous emissions, nevertheless, come into play in the presence of even far off-resonant lasers, which give rise to heating or loss.
The off-resonant heating rate and the effective Rabi frequency which
flips the pseudo spin in the Raman scheme scales
the same with respect to the ratio of laser power to detuning, and the maximum detuning while remaining spin sensitive is limited by the excited state fine structure splitting. Thus it is impossible to suppress heating
at a fixed Raman coupling, at least for alkali atoms.
This makes spontaneous heating effect much stronger for K than for Rb and Cs atoms.
The situations are worse for Na and Li atoms, whose relatively small fine structure splittings
essentially rule out the application of the Raman scheme.
To overcome this heating restriction, alternative schemes were proposed,
such as using narrow-line transitions of high spin atoms \cite{CuiPRA2013,DengPRL2012}
or manipulating spin-dependent tunneling without spin-flip in an optical lattice tilted by a static gradient magnetic field (GMF) \cite{KennedyPRL2013}.

Besides the restrictions on heating from atomic spontaneous emission,
the SOC strength realized with the Raman scheme is determined by the photon recoil momentum
and the intersection angle of the two Raman lasers. Thus it is difficult to tune
SOC strength continuously given a fixed
geometric setup in an experiment, although periodic modulation to
the effective Rabi frequency can be adopted to tune the SOC strength
smaller \cite{ZhangNat2013}, which was recently realized experimentally \cite{JimenezGarciarxiv2014}.

To overcome atomic spontaneous emission and tunability restrictions, one can seek out spin-dependent
interactions of atoms with magnetic fields to synthesize SOC, as was proposed by Xu {\it et al.} \cite{YouPRA2013} using repeated GMF pulses and by Anderson {\it et al.} \cite{SpielmanPRL2013-2} using modulating GMFs. A GMF provides a spin-dependent force, which over times leads to a spin (atomic internal state) dependent momentum (spatial/orbital degrees of freedom) impulse,
hence gives rise to SOC. An analogous protocol is also proposed to generate SOC for atoms in an optical lattice \cite{StruckPRA2014}. These ideas can be further extended to more general forms of SOC and in principle apply to all atoms with spin-dependent ground states. The procedure for obtaining an effective Hamiltonian from such a periodically driven quantum system is developed in Ref.\cite{GoldmanPRX2014}

\begin{figure}[htp]
\includegraphics[width=\columnwidth]{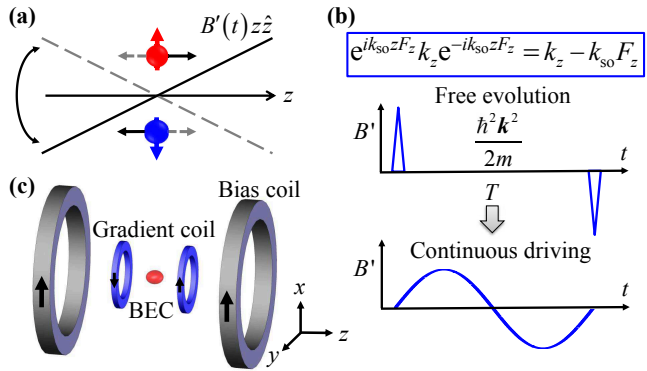}\\
\caption{A schematic illustration of SOC synthesized from a periodic
GMF.
(a) A periodically modulated GMF with zero
average $B'(t)z\hat{z}$ imparts opposite forces
(black arrows or gray dashed arrows at different times) to the
$|m_F=1\rangle$ (red disk and arrow) and $|m_F=-1\rangle$ (blue disk and arrow) states
of the $F=1$ hyperfine manifold. (b)
Each modulation period is composed of a pair of short opposite
GMF pulses (blue triangles), which provide impulses
$\pm \hbar k_{\rm so}$ with free evolution sandwiched in between,
translating canonical momentum $k_z\rightarrow k_z-k_{\rm so}F_z$,
leading to the SOC as shown in the blue rectangular box \cite{YouPRA2013}.
The continuous driving limit of a sinusoidal modulation with zero average is
adopted in our experiment for its better technical stability \cite{SpielmanPRL2013-2}.
(c) The experimental setup involves bias (gray) and gradient (blue) magnetic
coils. A BEC (red football) is placed at the center of the gradient coils
and aligned along the bias field.
} \label{Fig1}
\end{figure}

This Letter reports our experimental observation of
tunable SOC with equal Rashba and Dresselhaus contributions in a spin-1
$^{87}$Rb atom Bose-Einstein Condensate (BEC) synthesized by modulating
a one dimensional (1D) GMF \cite{YouPRA2013,SpielmanPRL2013-2}.
The effect of SOC is confirmed through the following two observations:
first, we observe the excitation of the collective dipole oscillation of
a condensate in a harmonic trap after abruptly turning on the SOC;
and second, we observe the adiabatic loading of condensed atoms
into the minimum energy state at shifted momentum when SOC is gradually
turned on by adiabatically ramping up
the amplitude of the modulating GMF.

We start with a brief review of the theoretical ideas \cite{YouPRA2013,SpielmanPRL2013-2}
for synthesizing SOC using GMF in reference to
the schematic illustration of Fig. \ref{Fig1}, where Fig. \ref{Fig1}(a) depicts
a temporally modulating GMF $B'(t)z\hat{z}$ with zero average
providing a spin-dependent force $g_F\mu_B
B'(t)F_z\hat{z}$ to a spin $F$ atom (with mass $m$) \cite{YouPRA2013}.
Here, $\mu_B$ is the Bohr magneton, $g_F$ is the
Lande g-factor and $F_{x,y,z}$
denotes the $x$-, $y$-, and $z$- component of spin vector $\textbf{F}$.
The origin for our SOC can be understood through
Fig. \ref{Fig1}(b), where an ultrashort GMF pulse
with impulse of $\hbar k_{\rm so}$ performs a spatial-dependent spin rotation
$U_z=\exp\{-ik_{\rm so}zF_z\}$. The two opposite GMF pulses then enact a unitary transformation,
which displaces the canonical momentum by a spin-dependent quantity,
{\it i.e.}, $U_z^\dag {k_z}{U_z} = {k_z} - {k_{{\rm{so}}}}{F_z}$.
Hence, the two pulse sequence is equivalent to an evolution with the effective Hamiltonian
$H_{\rm eff}=\hbar^2(k_z-k_{\rm so}F_z)^2/2m$ over a period $T$.
This analysis is consistent with a generalized protocol \cite{SpielmanPRL2013-2}
using a periodically modulated GMF with zero average
$\mathbf{B}(t)=\beta(t)(\hbar k_{\rm so}/g_F\mu_B)z\hat{z}$, where
$\beta(t)=\beta(t+T)$ and $\int_0^T{\beta(t)dt}=0$. The effective Hamiltonian
is modified to
\begin{eqnarray}
H_{\rm eff}=\frac{\hbar^{2}k^2_z}{2m}-\frac{c_1\hbar^{2}k_{\rm
so}}{m}k_zF_z+(\hbar q + \frac{c_2\hbar^{2}k^2_{\rm so}}{2m})F^2_z,
 \label{e1}
\end{eqnarray}
with $c_n=\int_0^{T}dt[\int_0^{t}\beta(t')\mathrm{d}t']^n/T$. $\hbar q$ is the quadratic Zeeman shift
of the bias field used for selecting the 1D GMF from a 3D quadrupole magnetic field (see supplemental material).
The second and the third terms on the rhs of Eq. (\ref{e1}) are the synthesized SOC
and an effective quadratic Zeeman shift (QZS), respectively.
The latter can be further tuned to zero or negative by an off-resonant dressing microwave field \cite{BlochPRA2006}.
In the experiment, we use a sinusoidal modulation function
$\beta(t)=\pi/T\sin(2\pi t/T)$, which gives $c_1=1/2$ and $c_2=3/8$ as in \cite{SpielmanPRL2013-2}.
By integrating the GMF in time, ${g_F}{\mu _B}\int_0^{T/2} {\beta \left( t \right)dt}  = \hbar {k_{{\rm{so}}}}$,
$\hbar k_{\rm so}$ is found to be equal to the momentum impulse from the GMF over half a modulation period.
The above physical picture remains approximately valid in the presence of an external trapping potential
provided the modulation frequency $\omega=2\pi/T$ is far greater than
the trapping frequency.
\begin{figure}
\includegraphics[width=\columnwidth]{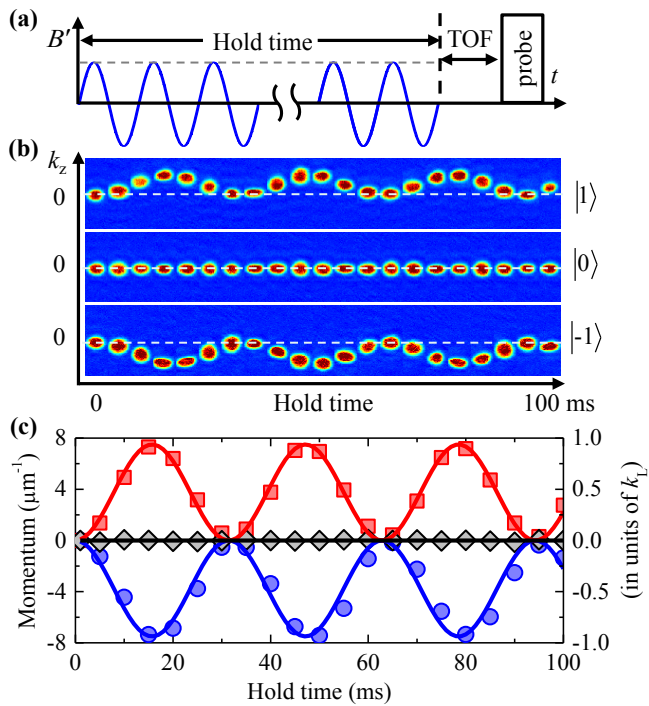}\\
\caption{Collective dipole oscillation of a single spin component condensate observed after abruptly
turning on $k_{\rm so}=7.5\, \mu m^{-1}$. (a) The time sequence of our experiments: abruptly turning on a constant amplitude modulation GMF corresponding to $k_{\rm so}=7.5\, \mu m^{-1}$, followed by 24ms of time of flight (TOF) before Stern-Gerlach imaging. The modulation period is $T=1\, ms$. (b) Absorption images of $|1\rangle$ (top row), $|0\rangle$ (middle row), and $|\!-\!1\rangle$ (bottom row) components after
different holding time (duration of the modulating GMF). The dashed lines are for $k_z=0$, or without SOC. (c) Atomic momentum for $|m_F=1\rangle$
(red squares), $|0\rangle$ (black rhombuses), and $|\!-1\!\rangle$ (blue disks) spin components as a
function of hold time. The rhs vertical labels are in units
of the resonant photon recoil momentum $k_L$ (at a wavelength of 780 nm), and solid lines denote theoretical predictions. } \label{Fig2}
\end{figure}

Our experiment is based on a single chamber BEC setup as described elsewhere \cite{GaoCPL2014}.
We create a $^{87}$Rb BEC of $1.2\times 10^5$ atoms in state $\left| {F=1,m_F=-1} \right\rangle$ in a crossed
dipole trap (Fig. \ref{Fig1}c) with trapping frequencies $(\omega_x,\omega_y,\omega_z)=2\pi \times (74.6,67.5,31.8)$ Hz. The 1D GMF $B'z\hat{z}$ is implemented by a combination of a 3D
quadrupole magnetic field $\textbf{B}_q=-B'x{\hat
x}/2-B'y{\hat y}/2+B'z{\hat z}$ and a 5.7 Gauss bias field
$\textbf{B}_b=B_0\hat{z}$, whose linear and quadratic Zeeman shifts are $(2\pi)\, 4$ MHz and $(2\pi)\, 2.34$ kHz, respectively (see supplemental material).
In addition to selecting the direction of the GMF,
the strong bias magnetic field $\textbf{B}_b$ suppresses magnetic field fluctuations from eddy currents.
The modulation frequency of the GMF $\omega$ is
$(2\pi)\, 1.0$ kHz unless stated otherwise.
The centers of the gradient field coils and the optical trap are aligned within $50\mu
$m to minimize short-term magnetic field fluctuation during
modulation. More details about magnetic field control can be found
in the supplemental material.

\begin{figure}
\includegraphics[width=\columnwidth]{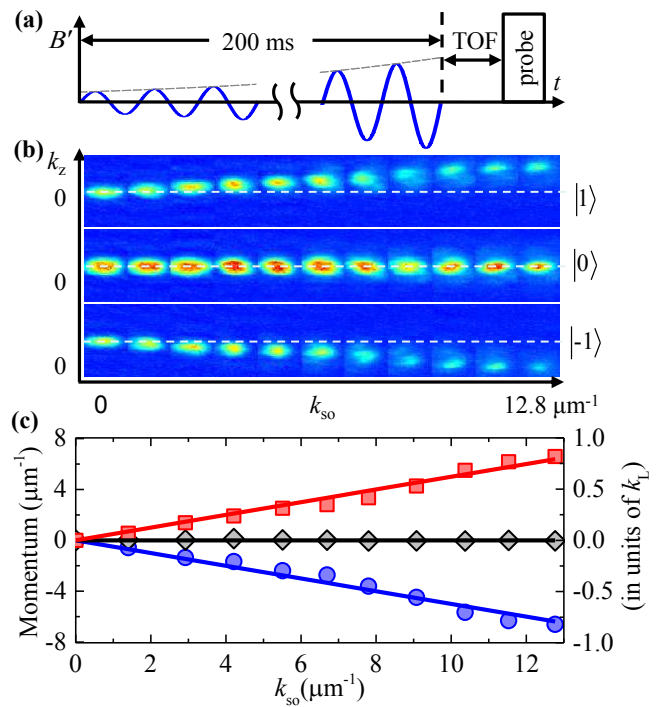}\\
\caption{Atoms adiabatically follow the energy minimum shifted to finite momentum with increased SOC strength. (a) The time sequence of our experiments: adiabatically ramping up GMF modulation amplitude to a given $k_{\rm so}$, followed by TOF before Stern-Gerlach imaging.
$T=1 ms$. (b) Absorption images of
$|1\rangle$ (top row), $|0\rangle$ (middle row), $|\!-1\!\rangle$ (bottom row) components
for different $k_{\rm so}$. The dashed lines are for $k_z=0$, or for without SOC. (c) Atomic center of mass momentum for $|1\rangle$ (red squares), $|0\rangle$ (black rhombuses), and $|\!-1\!\rangle$ (blue disks) spin component
as a function of $k_{\rm so}$, compared with theoretical predictions (solid lines).
} \label{Fig3}
\end{figure}

{\it Dipole oscillations} To confirm the effect of the synthesized SOC from the modulated GMF, we first excite the collective dipole oscillation of a single spin component atomic condensate in a harmonic trap by abruptly turning on the modulating GMF. By rewriting the effective Hamiltonian (\ref{e1}) as ${H_{{\rm{eff}}}} = {{{\hbar ^2}}}{\left( {{k_z} - {c_1}{k_{{\rm{so}}}}{F_z}} \right)^2}/{{(2m)}} + m\omega _z^2{z^2}/2$ and interchanging
the roles of the conjugate variable pair $k_z$ and $z$, it's easy to find that this effective
Hamiltonian is equivalent to that of a particle in a displaced harmonic trap,
where the extra QZS term only causes an overall energy shift and can be neglected.
A particle displaced from the
center of a harmonic trap will oscillate back and forth harmonically, which is indeed what we observe.
Both the position and momentum of the condensate oscillate at the trap frequency $\omega_z$.
Solving the Heisenberg equations of motion given by $H_{\rm eff}$, we obtain the averaged
kinetic momentum $\left<k_z\right>=c_1k_{\rm so}\left<F_z\right>[1-\cos(\omega_z t)]$, where $\left<F_z\right>=m_F$ for the $|m_F\rangle$ component. The oscillation is around $c_1k_{\rm so}m_F$
with a peak to peak amplitude $|2c_1k_{\rm so}m_F|$. In our experiments we abruptly turn on the modulating GMF, for instance as shown in Fig. \ref{Fig2}(a) at a corresponding SOC strength of $k_{\rm so}=7.5\, \mu m^{-1}$,
and persist for variable hold time. At integer multiple periods of the modulation,
the crossed dipole trap holding the condensate is turned off in less than $10$ $\mu$s.
Condensed atoms are expanded for about $24$ ms, during which different
Zeeman components are Stern-Gerlach separated by an
inhomogeneous magnetic field along ${x}$-direction.
For all three spin components, atomic center of mass momentums are derived
from their shifted positions along ${z}$-direction
with respect to their locations when SOC is absent.
As shown in Fig. \ref{Fig2}(c), the observed results are in good agreement with our theoretical predictions.

{\it SOC shifted minimum energy state}
As a second confirmation, we observe atom's minimum energy state adiabatically adjusted
to a finite none-zero momentum $k_z=c_1 k_{\rm so} m_F$ for state $|m_F\rangle$
when the modulation amplitude is slowly ramped up as in Fig. \ref{Fig3}(a).
In the presence of SOC we discuss, the minimum of the dispersion relation for spin state $|m_F\rangle$
is located at $c_1 k_{\rm so} m_F$. According to adiabatic theorem, if the ramping of
$k_{\rm{so}}$ is slow enough, atoms at the minimum will follow the ramp and stay at the
shifted  minimum.
In this set of experiments, atoms are prepared in the initial spin state of
$(1/2,1/\sqrt{2},1/2)^T$ by applying a $\pi/2$ pulse to the state $(0,0,1)^T$.
The modulation amplitude is then ramped up to a final value in $200$ ms.
After turning off the optical trap, we measure the momentum for each $m_F$ component.
Again, we find good agreement with theoretical predictions as shown in
Fig. \ref{Fig3}(c). To confirm adiabaticity, $k_{\rm so}$ is ramped up
from $0$ to $4.9$ $\mu m^{-1}$ in $100$ ms and then
back to $0$ in another $100$ ms. We find
atomic center of mass momentum
returns to $0$ without noticeable heating.
We also check the dependence of $c_1$ on the
modulation frequency $\omega$ and find that $c_1$ essentially remains a constant
as long as $\omega\gg\omega_z$.
\begin{figure}
\includegraphics[width=\columnwidth]{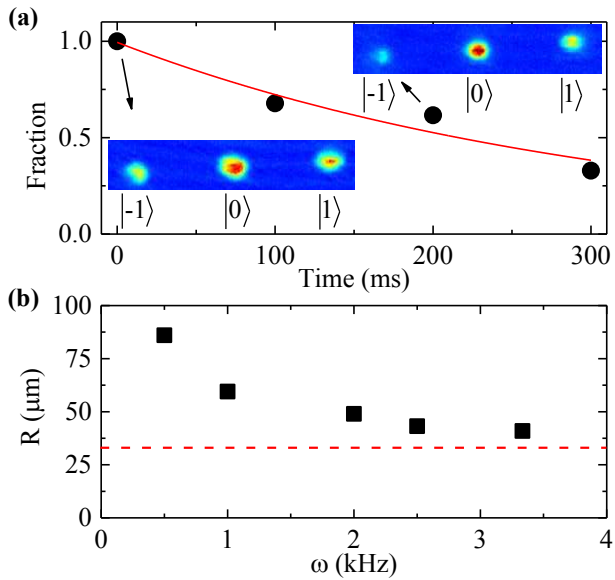}\\
\caption{(a) The fraction of remaining atoms (due to heating loss from the trap)
as a function of the time (after the modulating GMF is turned off)
 for $|m_F=-1\rangle$ component at $k_{\rm so}=4.9\, \mu m^{-1}$.
Dots denote measurement points, while the solid line is an exponential fit.
The optical trap frequencies are respectively $(\omega_x,\omega_y,\omega_z)=2\pi \times
(74.6,67.5,31.8)$ Hz, and $\omega=(2\pi)\,1.0$ kHz.
The insert Stern-Gerlach images show the three spin components
displaced along the vertical direction as a result of spin dependent momentum impulses from SOC;
(b) The condensate size $R$ (fitted radius) after 24 ms TOF expansion (black square) for
 $|\!-1\!\rangle$ component at $k_{\rm so}=4.9\, \mu m^{-1}$ as a function of
  modulation frequency $\omega$. The dashed line denotes the same
  size without SOC or at $k_{\rm so}=0\, \mu m^{-1}$. Hold time is 100 ms.
  The optical trap frequencies are respectively $(\omega_x,\omega_y,\omega_z)=2\pi \times
(60,100,60)$ Hz.
} \label{Fig4}
\end{figure}

The adiabatic result of Fig. \ref{Fig3} can be viewed
as demonstrating SOC tunability of the GMF scheme. In the following we briefly
discuss a plausible heating mechanism. Atomic spontaneous emission
is absent within the GMF scheme, we suspect the most likely heating mechanism comes
from parametric processes associated with temporal modulation.
To minimize parametric heating, we modulate far away
from the characteristic frequencies of our system:
the typical trap frequency at about $(2\pi)\,100$ Hz and
the mean field interaction energy at around $(2\pi)\,200$ Hz.
Modulating at $\omega=(2\pi)\,1.0$ kHz, heating is found to be moderate.
The worst case occurs for condensates in the $|-1\rangle$ state,
whose life time is found to be around $310$ ms based on fitting the
measured fractions of remaining atoms as a function of time
shown in Figure \ref{Fig4}(a) for atoms prepared in
the $(1/2,1/\sqrt{2},1/2)^T$ state and at $k_{\rm so}=4.9$ $\mu m^{-1}$.
This life time translates into comparable heating rate reported
for the Raman coupling scheme with Rb atoms \cite{SpielmanNat2011},
and can be improved
with increasing modulation frequency. Figure \ref{Fig4}(b)
displays the atomic cloud size after 24 ms of TOF expansion at different modulation
frequencies and GMF with the same $k_{\rm so}$. The fitted cloud radius
clearly decreases with increasing modulation frequency,
and heating can be greatly suppressed at even higher modulation frequencies.
Thus enhanced performances of the GMF scheme is expected
if our experiments can be repeated with atomic chip based setups,
which routinely provide higher GMF and faster modulations \cite{SchummNat2005,MachlufNat2013}.

As is demonstrated, the SOC synthesized from GMF enacts
spin-dependent momentum shift to the single atom dispersion curves, potentially leading to
curve crossings between different spin states. Inspired by the
interesting idea of Ref. \cite{StruckPRA2014}, we find that these crossings
can be tuned into avoided crossings when spin flip mechanism is introduced
 as elaborated in more detail in the supplemental material.

In conclusion, we experimentally demonstrate tunable SOC synthesized with a modulating GMF for a spin-1 $^{87}$Rb BEC. We tune the synthesized SOC strength with changing momentum impulse to an atom by the
GMF. The observed coherence time is reasonably long compared with Raman coupling scheme, pointing to promising future experimental opportunities. The GMF scheme we realize relies on spin-dependent Zeeman interactions, as such it is naturally extendable to higher spin atomic states,
like the spin-1 case we demonstrate here.
Further extensions of our experiment can lead to more
general forms of SOC including atoms in optical lattices \cite{StruckPRA2014}, and in higher spatial dimensions like the full Rashba or Dresselhaus forms of SOC \cite{YouPRA2013,SpielmanPRL2013-2}. It adds to the recent report
of spin-1 SOC \cite{Campbell2015} concatenated from
two pseudo-spin-1/2 subsystems with Raman laser couplings. Our experiment opens up a new avenue
in the promising pursuit of observing non-Abelian topological phenomenon in ultracold quantum gases.

This work is supported by MOST 2013CB922002 and
2013CB922004 of the National Key Basic Research Program of China,
and by NSFC (No.~91121005,
No.~11374176, No.~11404184, and No.~11474347).

\bibliography{ref}{}

\end{document}

\title{Supplementary material: Tunable spin-orbit coupling synthesized with a modulating gradient magnetic field}

\author{Xinyu Luo}
\affiliation{State Key Laboratory of Low Dimensional Quantum
Physics, Department of Physics, Tsinghua University, Beijing 100084,
China}

\author{Lingna Wu}
\affiliation{State Key Laboratory of Low Dimensional Quantum
Physics, Department of Physics, Tsinghua University, Beijing 100084,
China}

\author{Jiyao Chen}
\affiliation{State Key Laboratory of Low Dimensional Quantum
Physics, Department of Physics, Tsinghua University, Beijing 100084,
China}

\author{Qing Guan}
\affiliation{Beijing National Laboratory for Condensed Matter Physics, Institute of Physics, Chinese Academy of Sciences,
Beijing 100080, China.}

\author{Kuiyi Gao}
\affiliation{Beijing National Laboratory for Condensed Matter Physics, Institute of Physics, Chinese Academy of Sciences,
Beijing 100080, China.}

\author{Zhi-Fang Xu}
\affiliation{Department of Physics and Astronomy, University of Pittsburgh, Pittsburgh, Pennsylvania 15260, USA}

\author{L. You}
\affiliation{State Key Laboratory of Low Dimensional Quantum
Physics, Department of Physics, Tsinghua University, Beijing 100084,
China} \affiliation{Collaborative Innovation Center of Quantum
Matter, Beijing, China}

\author{Ruquan Wang}
\email{ruquanwang@iphy.ac.cn}
\affiliation{Beijing National Laboratory for Condensed Matter Physics, Institute of Physics, Chinese Academy of Sciences, Beijing 100080, China.} \affiliation{Collaborative Innovation Center
of Quantum Matter, Beijing, China}

\date{\today}
\pacs{67.85.De, 03.75.Mn, 67.85.Jk}
\maketitle

This supplemental material addresses a few interesting points in more detail.
\section{One dimensional (1D) gradient magnetic field (GMF)}
According Maxwell's equations both the divergence and curl of a static magnetic field
vanish in free space. Hence a one dimensional (1D) gradient magnetic field (GMF) $B'z{\hat z}$
cannot exist alone. In our experimental implementation, the $z$-gradient is
selected from a 3D quadruple field $-B'x{\hat x}/2-B'y{\hat y}/2+B'z{\hat z}$ by a strong bias field $B_0{\hat z}$, which corresponds to a linear Zeeman shift $\omega_0=g_F\mu_BB_0/\hbar$. Spin flips can be introduced by a radio-frequency (RF) transverse magnetic field $b\cos(\omega t){\hat x}$. In the frame rotating at $\omega$ and under rotating-wave approximation, the Hamiltonian containing the Zeeman shift of the bias field, the 3D quadrupole magnetic field, and the RF magnetic field reads
\begin{eqnarray}
\tilde{H}_B=(g_F\mu_BB'z-h\Delta) F_z + \hbar qF^2_z + \hbar\Omega F_x,
\label{s1}
\end{eqnarray}
provided $B_0\gg B'y/2,B'x/2$, where $\Delta=\omega-\omega_0$ is the detuning of the RF field frequency from the linear Zeeman shift, $|q|={\omega^2_0}/{\Delta_{\rm HFS}}$ is the quadratic Zeeman shift (QZS) of the bias field with $\Delta_{\rm HFS}$ the hyperfine splitting of ground state, and $\Omega=g_F\mu_Bb/2\hbar$ is the corresponding resonant Rabi frequency of the RF field. When $\Delta=0$ and $\Omega=0$ and neglecting QZS, this Hamiltonian reduces to $\tilde{H}_B=g_F\mu_BB'z F_z$, which describes the Zeeman interaction of an atomic hyperfine spin $\textbf{F}$ with an effective 1D
GMF $B'z{\hat z}$.

\section{The Effective Hamiltonian with transverse magnetic field pulses}
\begin{figure}[!htb]
\includegraphics[width=0.5\columnwidth]{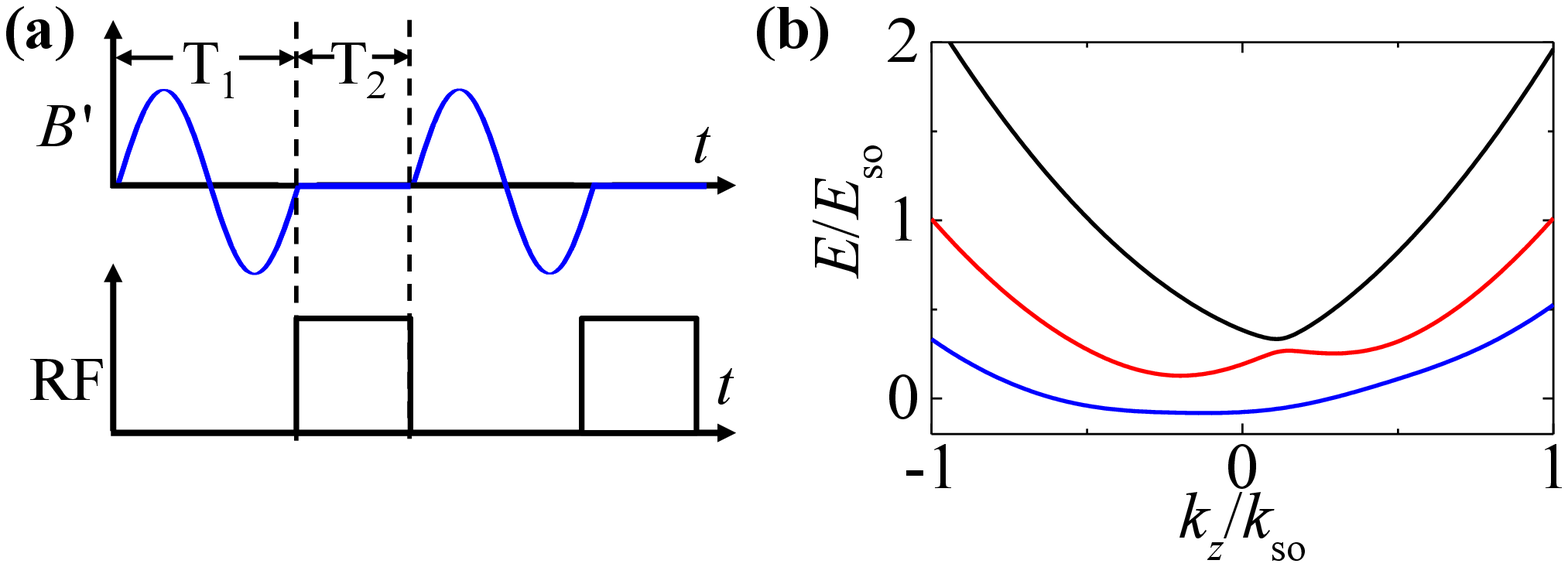}\\
\caption{Spin flip coupling synthesized from interlacing (a) sinusoidal modulations of the
GMF (blue wavy lines) of duration $T_1$ with radio-frequency pulses (black rectangles) of duration $T_2$. (b) The avoided crossing dispersion curves for $c'_1=0.4$, $\hbar\Omega'=0.15E_{\rm so}$, $\hbar\Delta=0.1E_{\rm so}$, $\hbar q'=0.25E_{\rm so}$.
} \label{Figs1}
\end{figure}
The present scheme we demonstrate gives rise to an effective
single atom SOC without explicit spin flip interaction.
Due to the presence of a large bias field, different spin states
are off-set by large energies, which hinders the effectiveness
of atom-atom interactions to access interesting many body quantum phases.
We can introduce spin flip interaction to our protocol
following an earlier suggestion in Ref. \cite{StruckPRA2014}.
In short, we augment our operational protocol by RF magnetic field pulses
during intervals between GMF modulation periods,
as shown in Fig. \ref{Figs1}(a), where $T_1$ and $T_2$ are respectively the durations of the GMF and the RF pulses. $T=T_1+T_2$ is the duration of one period. The evolution operator of one period is given by $U(T)=\exp\{-iH_{\rm so}T_1/\hbar\}\exp\{-iH_{\rm RF}T_2/\hbar\}$, where $H_{\rm so}={\hbar^{2}k^2_z}/({2m})-{c_1\hbar^{2}k_{\rm so}}k_zF_z/{m}+\hbar\Delta F_z+ (\hbar q+c_2E_{\rm so})F^2_z$, $H_{\rm RF}={\hbar^{2}k^2_z}/({2m})+\hbar\Omega F_x+\hbar\Delta F_z+ \hbar qF^2_z$, and $E_{\rm so}={\hbar^2k^2_{\rm so}}/{2m}$. When
the pulses are sufficiently short, $U(T)\approx\exp\{-i\left[H_{\rm so}T_1/T+H_{\rm RF}T_2/T\right]T/\hbar\}$. Thus the effective Hamiltonian approximates
\begin{eqnarray}
\begin{aligned}
H_{\rm eff}=&\frac{\hbar^{2}k^2_z}{2m}-\frac{\tilde{c}_1\hbar^{2}k_{\rm so}}{m}k_zF_z+\hbar \tilde{\Omega} F_x+\hbar\Delta F_z+\hbar \tilde{q}F^2_z,
\label{e2}
\end{aligned}
\end{eqnarray}
where $\tilde{c}_1=c_1{T_1}/{T}$, $\hbar \tilde{q}=\hbar q + c_2E_{\rm so}{T_1}/{T}$, and $\tilde{\Omega}=\Omega{T_2}/{T}$.
A gap now opens up between different eigenstate dispersions
as is shown in Fig. \ref{Figs1}(b).
This fine tuning is essential to the rich physics
already probed in previous SOC experiments \cite{SpielmanNat2011, PanPRL2012,ZhangPRL2012,MartinPRL2012,BeelerNat2013,ZhangNat2014,PanNat2014},
and constitutes one of the most important future directions for the present experiment.

\section{Condensate production}
Atomic $^{87}$Rb Bose-Einstein condensates are produced employing laser cooling
followed by evaporative cooling in a single vacuum chamber,
which consists of a 32 mm $\times$ 32 mm wide and 300 mm long quartz
cell, a Rb reservoir, and vacuum pumps. The quartz cell is ideally suited
for magnetic field modulation as it prevents eddy current, which could cause
short term fluctuations to the bias field. To optimize the applications of
fast and strong GMF modulation by small gradient coils
and allow for convenient upto three dimensional (3D) optical access,
MOT collected from background atoms are magnetically transported
10 cm to a location 8 mm away from the inside wall of the glass cell.
Transported atoms are first RF evaporatively
cooled in a hybrid trap consisting of a magnetic quadruple trap and
a crossed optical dipole trap for 5 seconds. The crossed dipole trap
derives from two 1064 nm laser beams (with $1/e^2$ Gaussian diameter 120
$\mu m$) propagating along $x$- and $z$- directions, respectively. The
magnetic quadrupole field gradient is then ramped down to transfer atoms
to the crossed dipole trap. Condensation occurs after a final evaporation
by ramping down the power of the crossed dipole trap in 3
seconds. At the end of the final evaporation the
crossed dipole trap frequencies are $(\omega_x,\omega_y,\omega_z)=2\pi \times
(60,100,60)$ Hz. A condensate in $|F=1, m_F=-1\rangle$ state with
$1.2\times 10^5$ atoms is produced every 40 seconds.
To minimize heating from the near-resonant driving of the
modulating GMF, we
further ramp down $\omega_z$ to $2\pi \times 31.8$ Hz in 500 ms
after BEC production, which reduces $\omega_x, \omega_y$ respectively to
$2\pi \times 74.6$ Hz and $2\pi \times 67.5$ Hz.

\section{Magnetic field control}
Three pairs of Helmholtz coils are used to control the homogeneous bias
magnetic field. While transferring atoms from the hybrid trap into
the crossed dipole trap, a $0.7$ Gauss bias field along $z$-direction
is simultaneously turned on in order to maintain atoms in
$|F=1, m_F=-1\rangle$ state. In the last 1.5
seconds of evaporative cooling, we ramp up the bias field to 5.7 Gauss
and hold on to this value. The Larmor frequency of the bias field is calibrated
by RF driven Rabi oscillations between Zeeman sublevels. The residual
magnetic field gradient is compensated to within 2 mGauss/cm by a pair
of anti-Helmholtz coils along $z$-direction.

A pair of small anti-Helmholtz coils is used to modulate the GMF
with a modulation amplitude up to 100
Gauss/cm at a frequency of $(2\pi)\,1.0$ kHz. The radius for the gradient coils is 15 mm.
The two coils are separated at an inner distance of 36 mm.
Each coil consists of 12 turns of winding and produces an inductance of about 10 $\mu
H$. The gradient coil size is much larger than the 120 $\mu m$ beam
waist of our dipole trap, which produces a homogeneous gradient field
inside the crossed dipole trap. The coils are small enough
to ensure fast and strong GMF modulation.
The current for the gradient coils is
regulated by a fast (10 $\mu s$ rise time) and precise (100 ppm) bipolar
current controller. The gradient coil is mounted on a 3D low
magnetic translation stage for precise alignment of
the gradient coils. The center of the gradient field is aligned within 50
$\mu m$ to the BEC, which is found to be crucial for minimizing short term
bias field fluctuations during GMF
modulation.

\bibliography{ref}{}